\documentclass{article}
\usepackage{spconf}
\usepackage{cite}
\usepackage{amsmath,amssymb,amsfonts}
\usepackage{algorithmic}
\usepackage{graphicx}
\usepackage{textcomp}
\usepackage{xcolor}
\def\BibTeX{{\rm B\kern-.05em{\sc i\kern-.025em b}\kern-.08em
    T\kern-.1667em\lower.7ex\hbox{E}\kern-.125emX}}

\usepackage{glossaries}
\usepackage{url}
\usepackage{hyperref}
\usepackage[caption=false,font=normalsize,labelfont=sf,textfont=sf]{subfig}
\newcommand{\E}[1]{\mathbb{E}\left(#1\right)}
\newcommand{\var}[1]{\text{Var}\left(#1\right)}

\newcommand{\vmin}{v_\text{min}(X)}
\newcommand{\vmax}{v_\text{max}(X)}

\title{A Mixed-Behavior Vote Model for Multimedia Subjective Quality Votes, Means, and Variances}
\name{Jaden Pieper and Stephen D. Voran}
\address{Institute for Telecommunication Sciences, Boulder, Colorado, USA \\
\{jpieper, svoran\}@ntia.gov}
\begin{document}

\ninept
\maketitle

\begin{abstract}
The relationship between subjective test vote variance and vote mean (or MOS) is well-studied, and the mathematically admissible vote variance region has been previously defined.
We propose a reduced admissible variance region called the Unimodal Variance Region (UVR) that better describes real subjective rating behavior of multimedia.
Further, subjective vote variance is often modeled as parabolic. 
We explain that, in practice, the parabolic model often violates the admissible region in the variance vs. MOS plane and we propose alternatives that respect the admissible region. 
We also present a parametrized random process to model votes that mixes voting processes and produces a realistic range of vote variances within the UVR at any desired MOS. 
This process was inspired by and comports with voting behavior that is observed in many subjective tests.
By modeling vote variance from a subjective experiment, this vote model offers additional interpretable insights into voting behavior observed in a given experiment.
We present example results from 16 datasets spanning speech, image, and video subjective quality experiments.

%\nw{Note that JP did not read this template closely --- do so before submitting anywhere if we use this format}
\end{abstract}


\section{Introduction}
Subjective tests are often used to evaluate various attributes of multimedia signals.
A very popular test format asks subjects to play a media file and then submit a vote to rate some pertinent attribute, often the overall quality, of each file by voting bad=1, poor=2, fair=3,  good=4, or excellent=5.
A mean opinion score (MOS) is found by averaging all votes for a given file. 
Specific testing protocols have been published for speech \cite{P.800,P.805,P.806,P.808}, audio \cite{BS.1116}, video \cite{BT.500,P.918,P.915, P.910}, multimedia \cite{P.910,P.920, P.1305}, and gaming  \cite{P.809} applications.

MOS values are the principal output of subjective tests, but the \emph{variance} of the votes for each file is also important since it describes the level of agreement for that file. A MOS of 1.0 or 5.0 requires all votes to agree, resulting in zero variance at the ends of the scale. 
The maximum and minimum possible values of vote variance for any MOS value, $1 \le X \le 5$, are given in \cite{Hossfeld2011} as
\begin{align}\label{eqn:maxVar}
    \vmax = (X -1)(5-X),  %\text{~and}
\end{align}
\begin{align}\label{eqn:minVar}
    \vmin = (X - \lfloor X\rfloor)(\lceil X \rceil - X),
\end{align}
where $\lfloor \cdot \rfloor$ and $\lceil \cdot \rceil$ denote the floor and ceiling operations. 
Note that $\vmax$ is produced when all votes for a file are 1s and 5s, and $\vmin$ is produced when all votes are adjacent (1s and 2s, for example).

The results in \cite{Hossfeld2011} apply only to the case of an infinite number of votes per file. 
We have extended that work to address the more practical situation of a finite number of votes per file, $2 \le n_v$. 
This extension requires scaling (\ref{eqn:maxVar}) or (\ref{eqn:minVar}) by a factor of $n_v/(n_v-1)$.  
% Examples are shown in Figure~\ref{fig:finite-nv-variance}. 
Together, these curves define the admissible range of variance values at any given values of $n_v$  and MOS. 
Throughout the rest of this paper we focus on the infinite vote case, for simplicity. 

Researchers have modeled the relationship between observed variance and MOS, $v_d(X)$, by scaling down the maximum variance parabola in (\ref{eqn:maxVar}) to best match the data\cite{Hossfeld2011, gao2025}.  
In other words, 
\begin{equation}\label{eqn:parabola-variance}
    v_d(X) = a(X-1)(5-X).
\end{equation}  
In \cite{Hossfeld2011} scale factors are found for 14 audio, video, and image datasets and they range from 0.038 to 0.249.
In \cite{gao2025} six datasets are fit, resulting in scale factors between 0.066 and 0.251.  
But values of $a < 0.25$ result in values of $X$ where $v_d(X) < \vmin$.  
That is, optimizing $a$ in (\ref{eqn:parabola-variance}) to match real data nearly always violates a fundamental limit that all data must obey.  
We conclude that the variance relationship in (\ref{eqn:parabola-variance}) can coarsely describe the main variance trends, 
but cannot fully describe real voting behavior.

The work in \cite{itsBounds} establishes bounds on subjective-to-objective correlation and mean squared error, and also provides a binomial-based subjective vote model called BinoVotes.  
The expected value of vote variance for BinoVotes intrinsically follows the curve
\begin{equation}\label{eqn:binovotes-variance}
    v_B(X) = 0.25(X-1)(5-x) = 0.25\cdot\vmax.
\end{equation}  
Since $0.25$ is the smallest scale factor that honors \mbox{$\vmin \leq v_d(X)$,} BinoVotes is a vote model that produces the smallest allowable parabolic variance relationship.

In the following, we propose a reduced admissible variance region that better describes observed subjective votes for multimedia (audio, video, images).
We also introduce a method of reshaping the maximum variance parabola (instead of simply scaling it) so that it better fits data without violating the minimum admissible variance.
Finally we build on BinoVotes to produce a vote model that can achieve any vote variance in the reduced variance region.
Together, the new method of fitting vote variance data with the new vote model yields interpretable insights into voting behavior for subjective experiments.

\section{Unimodal Variance Region}

Subjective tests collect votes and the first goal is to summarize all votes for each stimulus with a single statistic --- the mean. 
It is generally expected that the probability mass function (PMF) of the votes will be unimodal, meaning that there is a central consensus with some variability surrounding it depending on the stimulus and other experimental factors.
When the vote PMF is not unimodal, the mean is still a valid mathematical construct, but it fails to capture the most important characteristics of the votes.  

Multimodal PMFs indicate multiple groupings of distinct perceptions for a single stimulus.  With careful experimental design, this should be a very rare occurrence.
The maximum variance curve of (\ref{eqn:maxVar}) is achieved by extreme bimodal voting behavior where a stimulus receives only votes of 1 and 5.
This is a mathematical possibility, but if this pattern were observed consistently, it would raise methodological concerns and point to serious experimental design flaws or poor voter behavior.
We seek to define an admissible variance region that is consistent with more realistic voting behavior.

Using the same notation as~\cite{itsBounds}, let $Y$ denote a true quality, while $X$ represents MOS.
Further we only consider well-behaved vote models such that $\E{X|Y} = Y$, which is defined in more detail in~\cite{itsBounds}.
We now define the maximum variance unimodal (MVU) vote PMF and the adjacent two-choice (ATC) vote PMF, and propose the Unimodal Variance Region (UVR).
Building off of \cite{seaman_maximum_1985}, we define the MVU vote PMF when the stimulus has quality, $1 < y < 5$, as\footnote{PMFs for $y=1$ and $y=5$ are trivial and are not considered here.}
\begin{equation}\label{eqn:mvu-pmf}
    P(R=k|y, \text{MVU}) = \begin{cases}
        p_0, & 1 \leq k \leq y \\
        p_1, & y < k \leq 5 \\
        0, &  \text{otherwise},
    \end{cases}
\end{equation}
for $k=1,\hdots, 5$, where $R$ is the vote or rating.
The probabilities $p_0$ and $p_1$ satisfy $\E{R|y} = y$ and 
are specified in Table~\ref{tab:mvuvm-params}, along with the variance of the PMF.
Maximal variance is achieved by maximally spreading probability mass across the support. 
Simultaneous unimodality requires minimizing the number of unique probability values in the distribution. 
Achieving arbitrary means requires at least two unique probability values.
The result of these three constraints is (\ref{eqn:mvu-pmf}).

The ATC vote PMF is the PMF that achieves the minimum variance given in (\ref{eqn:minVar}). 
This PMF is unimodal, it has nonzero values for only two adjacent vote options as determined by the true quality,  $1 < y < 5$, and it satisfies $\E{R|y} = y$:
\begin{equation}
    P(R=k|y, \text{ATC}) = \begin{cases}
        \lfloor y \rfloor - y +1, & k =  \lfloor y \rfloor \\
        y - \lfloor y \rfloor, & k =  \lfloor y \rfloor + 1\\
        0, & \text{otherwise}.
    \end{cases}
\end{equation}

\begin{figure}
    \centering
    \includegraphics[width=\linewidth]{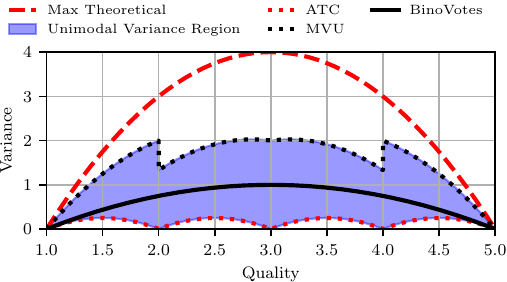}
    \caption{The Unimodal Variance Region and other variance relationships.}
    \label{fig:valid-variance}
\end{figure}

We use the MVU vote variance curve from (\ref{eqn:mvu-pmf}) and Table~\ref{tab:mvuvm-params} and the ATC vote variance curve given in (\ref{eqn:minVar}) to define the UVR, shown as the shaded region in Fig.~\ref{fig:valid-variance}.

\begin{table}[]
    \centering
    \caption{Parameters and variance for the MVU vote PMF.}
    \begin{tabular}{cccc} \hline
        \textbf{Domain} & $\mathbf{p_0}$ & $\mathbf{p_1}$ & $\mathbf{v_M(y)}$  \\ \hline
         $1 \leq y < 2$  & $(7 - 2y)/5$ &  $(y - 1)/10$ & $(y-1)(4-y)$ \\ \hline
         $2 \leq y < 3$ & $(4 - y)/ 5$ & $(2y - 3)/15$ & $-y^2 + 17y/3 - 6$  \\ \hline
         $3 \leq y < 4$ & $(9 - 2y) / 15 $ & $(y - 2)/5$ & $-y^2 + 19y/3 - 8$  \\ \hline
         $4 \leq y < 5$  & $ (5 - y)/10 $ & $(2y - 5) / 5$ & $(y -2)(5 - y)$  \\ \hline
    \end{tabular}
    \label{tab:mvuvm-params}
\end{table}

%% Note that BinoVotes is not the maximum variance model but we do not consider that model because we don't see the need for it in the data and it is not realistic.

\section{Modeling Vote Variance}

% \nw{In parabola scaling world then $0.05 \leq a \leq 0.54$. $a < 0.05$ means on average data is below minimum-variance curve (impossible) and $a> 0.54$ means on average vote behavior is not unimodal (undesirable).}

% Desireable properties of a vote variance function
%% 0 at edges of scale.
%% Symmetric about middle of the scale.
%% Continuous and smooth (continuous 1st derivative).
% This leaves us with the need to define a valid target variance function.
Next we define a parametrized variance function that can accurately model observed data while respecting the UVR.
In order to properly characterize subjective voting behavior, the function must have five desirable traits:
(1) take a value of 0 at both extremes of the scale;
(2) show symmetry about the middle of the scale;
(3) be continuous and smooth (continuous 1st derivative);
(4) be within the UVR (shaded region in Fig.~\ref{fig:valid-variance});
and (5) can properly describe observed data.
The symmetry trait is consistent with the symmetry of the voting scale. 
We have studied a significant amount of subjective data and when a lack of symmetry is seen it can be traced to missing or limited data.
% symmetry justification - while data is not always symmetrically distributed in subjective tests, that would be a goal if one had no functional constraints on how much testing they could do.

We propose modeling vote variance via a fourth-degree polynomial of the following form:
\begin{equation}\label{eqn:fourth-degree-poly-variance}
    v_r(Y) = (Y - 1)(5 - Y)(w_1(Y - 3)^2 + w_0).
\end{equation}
This can be interpreted as scaling $\vmax$, not by a single constant, but by a second parabola that attenuates the first one.
This yields variance functions that are flatter in the middle of the scale, which is often observed in data.
Given a dataset with $N$ MOS values, $X_i$, and corresponding vote variances, $\sigma^2_i$, we use least squares to find $w_0$ and $w_1$:
\begin{align}\label{eqn:weighted-least-squares}
   \min_{w_0, w_1} \! \sum_{i=1}^N \! \left( (\! X_i \! - \! 1)(5 \! - \! X_i)(w_1 (X_i \! - \! 3)^2 \! + w_0) \!  - \!  \sigma^2_i \right)^2 \! .
\end{align}

The variance function in (\ref{eqn:fourth-degree-poly-variance}) naturally enforces zero variance at the ends of the quality scale and symmetry about the middle of the quality scale, and it is continuous and smooth.
The least-squares solution for $w_0$ and $w_1$ allows $v_r(Y)$ to properly describe real data, but can also result in occasional small violations of $\vmin$ at the ends of the quality scale.
In these cases this is efficiently solved by using weighted least squares to assign small additional weight to MOS values near 1 and 5.
% This is efficiently solved by assigning small additional weight ($1 < \lambda_i$) to MOS values near one and five.
%By minimizing this weighing we can obtain curves that describe the data well (they have low overall fitting residuals) and they also stay inside the UVR. 
The introduction of these small weights causes only very minor changes in the variance functions, it maintains low overall fitting residuals, and eliminates the minor excursions outside of the UVR (below $\vmin$) that occur when fitting some data.
We have written software that iterates to efficiently find an ideal weighting when necessary and that software is available at \mbox{\url{https://www.github.com/NTIA/ITS-MOS-Agreement}}.
%In practice we have observed that unweighted least squares yields valid variance curves, and when weighting is required the resulting curve is typically visually indistinguishable from the unweighted version.

\section{Mixed-Behavior Vote Model}
The binomial-based vote model BinoVotes~\cite{itsBounds} is highly intuitive and has many desirable properties.
Foremost among these is that it respects vote characteristics --- votes are discrete and lie on the voting scale.
It is also well-behaved, meaning that the expected value of the votes is the true quality.
This naturally leads to simulated MOS values that respect real MOS characteristics.
The vote PMFs from BinoVotes are always unimodal and the vote probabilities decay as one moves away from the mode --- this follows our intuition for vote distributions.
Finally, BinoVotes assigns a non-zero probability to each of the five vote options, and we call it a five-choice vote model.
Specifically for $k=1,2,\hdots, 5$ the BinoVotes PMF is given by
\begin{align}
    % P(R=k|y, \text{BV}) = {4 \choose k-1} p^k (1-p)^k
    \!P(R=k|y, \text{BV}) \!=\! {4 \choose k-1}\! \left(\frac{y-1}{4}\right)^{k-1} \left(\frac{5-y}{4}\right)^{k-1}.
\end{align}

BinoVotes is driven by a single parameter --- the quality of the item being rated.
This allows BinoVotes to produce the desired MOS value, but BinoVotes cannot directly model voting behavior from specific subjective experiments, and thus cannot model the associated specific vote variances.
We have already noted that BinoVotes produces the smallest allowable parabolic variance relationship, corresponding to a value of $a=0.25$ in (\ref{eqn:parabola-variance}).
And Fig.~\ref{fig:valid-variance} shows that variance function sits roughly in the middle of the UVR.
These properties suggest that BinoVotes could be a good foundation for building a more flexible vote model.

Like the BinoVotes PMF, the MVU vote PMF and the ATC vote PMF that bound the UVR are both unimodal and have a single parameter.
The voting behavior for each is highly interpretable.
To model votes throughout the UVR, we propose to mix the BinoVotes PMF with either the ATC PMF or the MVU PMF.  
The result is the mixed-BinoVotes (MBV) PMF:
\begin{align}\label{eqn:mixed-pmf}
  \!\! \!\!\! P(R=k|y,\text{MBV})\! =\! \begin{cases}
        \alpha b_k + (1\! -\alpha)a_k, \!\!\!\!\! & v_r(y) \! \leq \! v_B(y) \\
        \alpha b_k + (1\! - \alpha) m_k,\!\!\!\!\! & v_r(y) \! > \! v_B(y).
    \end{cases}
\end{align}
Here $0 \leq \alpha \leq 1$ is a mixing parameter, $v_r(y)$ is the target variance for a given quality,  $v_B(y)$ is the variance produced by BinoVotes, $b_k = P(B=k|y)$, $a_k = P(A=k|y)$, and $m_k = P(M=k|y)$, and $B, A,$ and $M$ are random variables distributed as $B\sim\text{BinoVotes}$, $A\sim\text{ATC}$, and $M\sim\text{MVU}$.

All three PMFs used in MBV satisfy $\E{R|y}=y$, so the MBV PMF also satisfies
\begin{align}\label{eqn:mixed-mean}
    \E{R|y} 
    % &= \sum_{k} k P(R=k) \nonumber \\
    % &= \sum_k k(\alpha b_k + (1-\alpha) a_k \nonumber \\
    % &= \alpha \sum_k k b_k + (1-\alpha) \sum_k k a_k \nonumber \\
    % &= \alpha \E{B|y} + (1-\alpha)\E{A|y} \nonumber \\
    % &= \alpha y + (1-\alpha)y \nonumber \\
    &= y.
\end{align}
% The second moment of the mixture PMF is
% \begin{align}\label{eqn:mixed-second-moment}
%     \E{R^2|y} 
%     % &= \sum_{k} k^2 P(R=k) \nonumber \\
%     % &= \sum_k k^2(\alpha b_k + (1-\alpha) a_k \nonumber \\
%     % &= \alpha \sum_k k^2 b_k + (1-\alpha) \sum_k k^2 a_k \nonumber \\
%     &= \alpha \E{B^2|y} + (1-\alpha)\E{A^2|y}.
% \end{align}
% From (\ref{eqn:mixed-mean}) and (\ref{eqn:mixed-second-moment}) it follows that the variance of the mixture PMF is
It is easy to show that the variance of the mixture PMF is
\begin{align}\label{eqn:mixed-variance}
    \var{R|y} 
    % &= \E{R^2 | y} + \E{R|y}^2 \nonumber \\
    % &= \alpha \E{B^2 | y} + (1-\alpha) \E{A^2|y} - y^2 \nonumber \\
    &= \alpha(y) v_B(y) + (1-\alpha(y))v_A(y),
\end{align}
where $v_A(y) = v_\text{min}(y)$.
Equation (\ref{eqn:mixed-variance}) holds with $M$ in place of $A$, with $v_M(y)$ as defined in Table~\ref{tab:mvuvm-params}.
It follows that for any target variance function $v_r(y)$ inside the UVR, we can find a mixing function that achieves that target:
\begin{align}\label{eqn:mixing_parmameter}
\alpha(y) = 
 \begin{cases}
        \frac{v_r(y)-v_A(y)}{v_B(y)-v_A(y)}, & v_A(y) \leq v_r(y) < v_B(y) \\[5pt]
        \frac{v_r(y)-v_M(y)}{v_B(y)-v_M(y)}, & v_B(y) \leq v_r(y) \leq v_M(y). 
        % \\[5pt]
        % 0, &v_r(y) > v_M(y).
\end{cases}
\end{align}

The MBV PMF provides a mixed-behavior vote model because the mixture parameter, $\alpha(y)$, mixes the voting behavior of a middle-ground vote model (BinoVotes) with the voting behavior of a more extreme vote model (ATC or MVU).
These mixing values can be interpreted as the degree to which voters followed different voting behaviors, as seen in the next section. % in a real subjective experiment.

\section{Application, Analysis, and Discussion}
\begin{figure*}[t]
\begin{center}
    \subfloat[]{
        \includegraphics[width=0.33\linewidth]{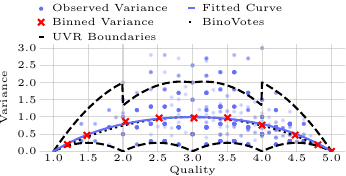}
        % \label{fig:fit-its2013}
        }
    \subfloat[]{
        \includegraphics[width=0.33\linewidth]{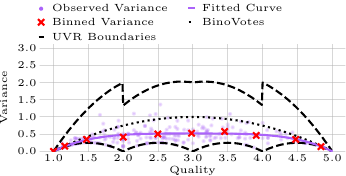}
        % \label{fig:fit-nisqa-p501}
        }
    \subfloat[]{
        \includegraphics[width=0.33\linewidth]{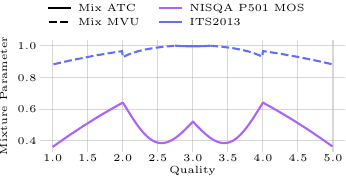}
        % \label{fig:mixture-param}
    }
    \end{center}
    \vspace{-7mm}
    \caption{Raw and averaged variance data for ITS2013 (left) and NISQA P501 MOS (middle) votes, the fourth-degree fitting (\ref{eqn:fourth-degree-poly-variance}) and other variance relationships as indicated. One outlier variance value above 3 not shown for ITS2013. Mixture parameters are shown in right panel.}
    \label{fig:data-fits}
\end{figure*}

\begin{figure}[h]
    \centering
    \includegraphics[width=\linewidth]{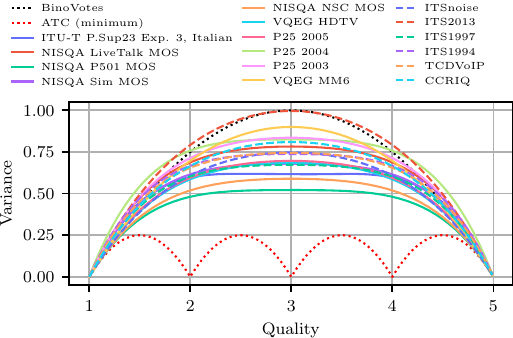}
    \caption{Fourth-degree vote variance fits (\ref{eqn:fourth-degree-poly-variance}) for 16 subjective tests.}
    \label{fig:fitted-variances}
\end{figure}
We modeled vote variance using (\ref{eqn:fourth-degree-poly-variance}) for 16 datasets, listed in Table~\ref{tab:datasets-parameters} along with the resulting values of  $w_0$ and $w_1$.
The resulting fitted curves can be seen in Fig.~\ref{fig:fitted-variances}.
Unweighted least squares gives results inside the UVR for 13 of the 16 datasets, and weighted least squares is needed to obtain results within the UVR for those final three datasets.

Figure~\ref{fig:data-fits} provides details for two datasets: ITS2013 and NISQA P501 MOS. It shows the raw vote variance values, the mean variance for 10 MOS bins, and the fitted curve (\ref{eqn:fourth-degree-poly-variance}). 
These two datasets have the largest and smallest mean variance curves of the 16 datasets considered and the fitted curve tracks these very closely in each case, while always remaining within UVR.
The associated mixture parameter values are shown in Fig.~\ref{fig:data-fits} (right panel). 

Figure~\ref{fig:data-fits} (left panel) shows that the ITS2013 variance curve matches the BinoVotes variance curve very closely and this is consistent with mixture parameter values near 1.0 seen in Fig.~\ref{fig:data-fits} (right panel).
On the other hand, the NISQA P501 MOS dataset has much smaller vote variance across the quality scale.
In fact, 28\% of the vote variance values (67 of 240 files) lie on the minimum variance curve.
This means that \emph{all} of the 18 to 32 votes received by each file (28.3 votes on average) were \emph{adjacent values only}.
This is consistent with the much lower mixture parameter values for NISQA P501 MOS in Fig.~\ref{fig:data-fits} (middle panel)
and this demonstrates that ATC does occur in subjective tests.
\begin{table}[]
    \centering
    \caption{
    Datasets and fitting parameters for the parabola in (\ref{eqn:parabola-variance}) and the fitting in (\ref{eqn:fourth-degree-poly-variance}).
    \textdagger~indicates image quality tests;
    \textdagger\textdagger~indicates video quality tests; all others are speech quality tests.
    *~Indicates tests of the land-mobile radio system in
    \cite{p25-subjective-tests-apco}.
    }
    \resizebox{\linewidth}{!}{
        \begin{tabular}{lrlrrr}
\textbf{Dataset} & \textbf{Files} & $\mathbf{n_v}$ & $\mathbf{a}$ & $\mathbf{w_0}$ & $\mathbf{w_1}$ \\ \hline
ITU-T P.Sup23 Exp. 3, Italian \cite{Psup23} & 200 & 24.0 & 0.173 & 0.154 & 0.045 \\ \hline
NISQA LiveTalk MOS \cite{Mittag2021IS} & 232 & 24.0 & 0.211 & 0.196 & 0.033 \\ \hline
NISQA P501 MOS \cite{Mittag2021IS} & 240 & 28.3 & 0.145 & 0.130 & 0.030 \\ \hline
NISQA Sim MOS \cite{Mittag2021IS} & 12500 & 5.2 & 0.188 & 0.170 & 0.036 \\ \hline
NISQA NSC MOS \cite{Mittag2021IS} & 240 & 27.2 & 0.160 & 0.147 & 0.026 \\ \hline
VQEG HDTV \cite{VQEGHDTVreport} \textdagger\textdagger & 1008 & 24.0 & 0.188 & 0.171 & 0.024 \\ \hline
P25 2005* & 1600 & 5.2 & 0.186 & 0.174 & 0.024 \\ \hline
P25 2004* & 1024 & 7.5 & 0.227 & 0.206 & 0.046 \\ \hline
P25 2003* & 1024 & 7.9 & 0.219 & 0.209 & 0.029 \\ \hline
VQEG MM6 \cite{PinsonMM6} \textdagger\textdagger & 600 & 21.3 & 0.230 & 0.225 & 0.007 \\ \hline
ITSnoise \cite{ITSimageNoise} \textdagger\textdagger & 288 & 20.8 & 0.205 & 0.187 & 0.016 \\ \hline
ITS2013 \cite{VoranICASSP2013} & 1180 & 5.6 & 0.255 & 0.249 & 0.015 \\ \hline
ITS1997 (Tests 16 and 17 in \cite{VoranSCW99}) & 720 & 6.4 & 0.183 & 0.168 & 0.033 \\ \hline
ITS1994 (Test 4 in \cite{MNBpartII}) & 2432 & 8.0 & 0.199 & 0.185 & 0.038 \\ \hline
TCDVoIP \cite{Harte2015} & 384 & 24.0 & 0.204 & 0.186 & 0.032 \\ \hline
CCRIQ \cite{CCRIQ} \textdagger & 2352 & 9.8 & 0.213 & 0.202 & 0.019 \\ \hline
\end{tabular}
    }
    \label{tab:datasets-parameters}
\end{table}
\begin{figure}
    \centering
    \includegraphics[width=\linewidth]{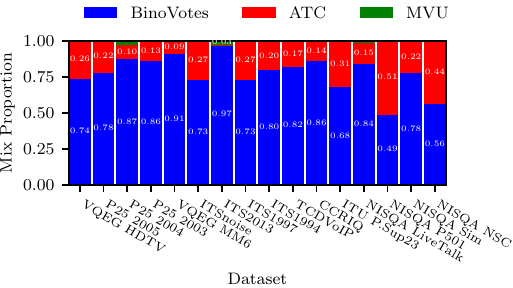}
    \caption{Summary of mixing parameter values for 16 datasets.}\vspace{-2mm}
    \label{fig:avg-mix}
\end{figure}

For any dataset, we can model observed vote variances using (\ref{eqn:weighted-least-squares}), use the model parameters $w_0$ and $w_1$ in (\ref{eqn:fourth-degree-poly-variance}) to get $v_r(y)$, and then use (\ref{eqn:mixing_parmameter}) to find the mixing parameter $\alpha(y)$.
This mixing function can be evaluated uniformly across the range $1 \leq y \leq 5$ to find the relative contributions of the three vote models (BinoVotes, ATC, MVU) to the subjective data.
Alternately, the mixing function can be evaluated at the MOS value of every file in the database to obtain a more data-driven view of the relative contributions of the three vote models.
The second approach yields the results in Fig.~\ref{fig:avg-mix}.
which gives a coarse view of voting behavior within each test, and shows how voting behavior differs between tests. 
For example, the ITS2013 data can almost entirely be described by BinoVotes voting behavior, while the NISQA P501 MOS data is described as a nearly equal mix of BinoVotes and ATC voting behaviors.

Vote variance has previously been modeled by the scaled parabola given in (\ref{eqn:parabola-variance}).
Applying this to the boundaries of the UVR can give insights into subjective voting behavior.
% The parabola has been well-studied, but without much attention to the implications of the value $a$ in (\ref{eqn:parabola-variance}).
Values of $a < 0.25$ cause the parabola to fall below the minimum possible variance curve (\ref{eqn:minVar}) at some MOS values, which is undesirable.
A more forgiving lower limit on $a$ can be found by fitting the parabola (\ref{eqn:parabola-variance}) to the minimum variance curve (\ref{eqn:minVar}).
The result is $a = 0.05$ and this can be interpreted as the smallest parabola scaling that describes observed subjective vote data which do not ``consistently'' violate the minimum variance curve.
%So any value of $a$ lower than this would , and that any values of $a$ lower than this would have to come from data that on average violates the minimum variance curve.
So, data that yield $a < 0.05$ should be examined to validate its integrity.
Similarly, fitting the parabola (\ref{eqn:parabola-variance}) to the MVU variance curve given in Table \ref{tab:mvuvm-params} yields a value of $a=0.54$.
So if fitting (\ref{eqn:parabola-variance}) to data gives $a > 0.54$, we would conclude that multimodal PMFs play a significant role in this data, and this could indicate issues with subject behavior, experimental design, or both.
Examples of values of $a < 0.05$ and $a > 0.54$ can be observed in the literature~\cite{Hossfeld2011}.

\section{Conclusion}
We have developed the UVR by replacing the mathematically possible but unrealistic maximum variance curve with the variance curve of the MVU vote model.
This upper bound describes the highest variance voting behavior that could be observed in a subjective experiment without pointing to improper subject rating behavior or experimental design flaws.
We have also proposed an improved methodology for modeling vote variance data that respects the UVR.
Lastly we have proposed a mixed-behavior vote model that can be used to quantify three types of voting behavior that may be observed in subject test results.
These two modeling steps can be used independently or in tandem to gain insights into subjective test vote variance and behaviors that can produce that variance.

\bibliographystyle{IEEEbibDOI}
%\bibliography{strings,refs}
\bibliography{sources}

\end{document}